\documentclass[%
 reprint,
superscriptaddress,
 amsmath,amssymb,
 aps,
 prl,
]{revtex4-2}

\usepackage{graphicx}
\usepackage{dcolumn}
\usepackage{braket}
\usepackage{bm}
\usepackage[colorlinks=True]{hyperref}

\begin{document}
\preprint{APS/123-QED}

\title{Time dynamics of multi-photon scattering in a two-level mixer}

\author{A.V. Vasenin}%
\affiliation{Skolkovo Institute of Science and Technology, Nobel St. 3, 143026 Moscow, Russia}
\affiliation{Laboratory of Artificial Quantum Systems, Moscow Institute of Physics and Technology, 141700 Dolgoprudny, Russia}
\author{A.Yu. Dmitriev}
 \affiliation{Laboratory of Artificial Quantum Systems, Moscow Institute of Physics and Technology, 141700 Dolgoprudny, Russia}
\author{S.V. Kadyrmetov}
\affiliation{Laboratory of Artificial Quantum Systems, Moscow Institute of Physics and Technology, 141700 Dolgoprudny, Russia}
\author{A.N.Bolgar}
\affiliation{Laboratory of Artificial Quantum Systems, Moscow Institute of Physics and Technology, 141700 Dolgoprudny, Russia}
\author{O.V. Astafiev}%
\affiliation{Skolkovo Institute of Science and Technology, Nobel St. 3, 143026 Moscow, Russia}
\affiliation{Laboratory of Artificial Quantum Systems, Moscow Institute of Physics and Technology, 141700 Dolgoprudny, Russia}
\affiliation{Physics Department, Royal Holloway, University of London, Egham, Surrey TW20 0EX, United Kingdom}
\date{\today}

\begin{abstract}

A superconducting qubit in a waveguide behaves as a point-like nonlinear element. 
If irradiated with nearly resonant microwave pulses, the qubit undergoes quantum evolution and generates coherent fields at sideband frequencies due to elastic scattering. This effect is called Quantum Wave Mixing (QWM), and the number of emerged side components depends on the number of interacting photons. By driving a superconducting qubit with short pulses with alternating carrier frequencies, we control the maximal number of photons simultaneously interacting with a two-level system by varying the number and duration of applied pulses. Increasing the number of pulses results in consecutive growth of the order of non-linearity, which manifests in additional coherent side peaks appearing in the spectrum of scattered radiation while the whole spectrum maintains its asymmetry. 

\end{abstract}

\maketitle



The scattering of electromagnetic waves on a single atom in an open space is a cornerstone problem in quantum optics \cite{cohen1998atom, Meystre1990, Walls1994}. 
A two-level system driven by resonant monochromatic tone is a great playground to create and study non-trivial light with sophisticated properties. In addition to the predicted \cite{Mollow1969,Kimble1976} and later observed \cite{Wu1975, Muller2007, Astafiev2010} intensity-dependent Rayleigh scattering and three-peaked inelastic spectrum, the scattered field exhibits direction- and power-dependent bunching \cite{Apanasevich_1979} and antibunching \cite{Kimble1977, Matthiesen2012, Phillips2020, Hanschke2020}, squeezing \cite{Walls1981,Vogel,Collett_1984,schulte2015quadrature}, sub-poissonian photon statistics \cite{Short1983} and spectral correlations \cite{Aspect1980, Nienhuis1993}, quantum amplification of probe signal \cite{Wen2019}. Therefore, various aspects of resonance fluorescence are well studied, finding its applications in microwave photonics and quantum information processing platforms based on propagating fields. However, altering the drive to a pair of tones (so-called bichromatic drive) complicates the stationary and dynamic characterization of the field emitted by dressed two-level system.

The pioneering experiments with the use of atomic vapours \cite{Zhu1990}, quantum dots \cite{Peiris2014} and superconducting qubits \cite{Pan2021} have revealed that an inelastic fluorescence spectrum under bichromatic drive becomes qualitatively different from the well-known Mollow triplet for the monochromatic drive. In the case of symmetrically detuned drives (that is, located at $\omega_{\pm} = \omega_{d}\pm \delta\omega$, where $\omega_d$ is central frequency of drives, typically equal to qubit's frequency $\omega_q$, and $\delta\omega$ is arbitrary chosen detuning) with equal Rabi amplitudes $\Omega_-= \Omega_+=\Omega$, the spectrum consists of many peaks. For any integer $n$ there is a peak at $\omega_{\pm n} = \omega_{d}\pm n \delta\omega$, and these frequencies do not depend on the $\Omega$ of each drive but their intensities do. This effect was explained with the direct Bloch equation solutions \cite{Tewari1990, Ficek1993}, in some works also with the use of dressed atom picture \cite{Agarwal1991,Freedhoff1990}, and the elaborated theory gives correct peak positions and heights. However, the \textit{elastic} components predicted \cite{Agarwal1991,Ruyten1992} to appear at $\omega_{\pm (2p+1)}$ for all integers $p$, were hardly observed in traditional quantum optics, partially because coherent field measurements are rather cumbersome in optics of visible range \cite{Leonhardt1993} when compared with, for example, photocounting measurements. The only observation of elastic scattering known to us was made with the use of high-finesse Fabry-Perot cavity \cite{Peiris2014}, however, the method is not phase-sensitive and intensities of elastic components for different parameters were not analyzed.  In contrary, the elastic field components are straightforward to observe in microwave scattering by a single superconducting qubit in a waveguide \cite{Astafiev2010,Abdumalikov2011, lu2021characterizing}. Their emergence is analogous to the coherent optical wave mixing in non-linear media. However, the medium consists of the only two-level system. Thereby the observed mixing is called the Quantum Wave Mixing (QWM) \cite{Dmitriev2017, Dmitriev2019, Teresa2020}, and it is efficient in the regime $\delta\omega \ll \Gamma_1$ and $\Omega_{\pm}\ge\Gamma_1$.
Moreover, it was proposed \cite{Dmitriev2019} and theoretically confirmed \cite{Pogosov2021} that QWM could reveal photon statistics of non-classical light in one of the modes. It could become a handy tool for microwave quantum optics: the absence of reliable photon detectors of sufficiently wide bandwidth and high efficiency is significant restriction for microwave waveguide photonics.

In current work, we measure a complete picture of QWM spectrum due to elastic scattering of microwave pulses when the qubit undergoes coherent dynamics. To achieve that, the qubit is driven by a series of $N$ non-overlapping pulses with alternating frequencies $\omega_-$~and $\omega_+$, with $N$ from 2 to 6. In the elastic spectrum, we observe solely $2N-1$ components which depend on the maximal number of interacting photons. Particularly, we extend the result of Ref. \cite{Dmitriev2017}, where two pulses (N = 2) represented maximum three photon interaction and, therefore, a quantized spectrum consists of three peaks only. We also measure the maps of complex amplitudes depending on durations of blue- and red-shifted pulses and fit them with analytical calculations, finding excellent agreement with measurements. In addition, we numerically find the evolution and characterize the dependence of side peaks for the specific case of partially overlapping three pulses. Our results demonstrate high degree of control of non-linear light scattering on a single atom in the time domain. They will serve as a basis for the development of non-linear quantum electronics with superconducting systems and for studying its transient properties under strong drive.

\begin{figure}[b]
\includegraphics[width=1\linewidth]{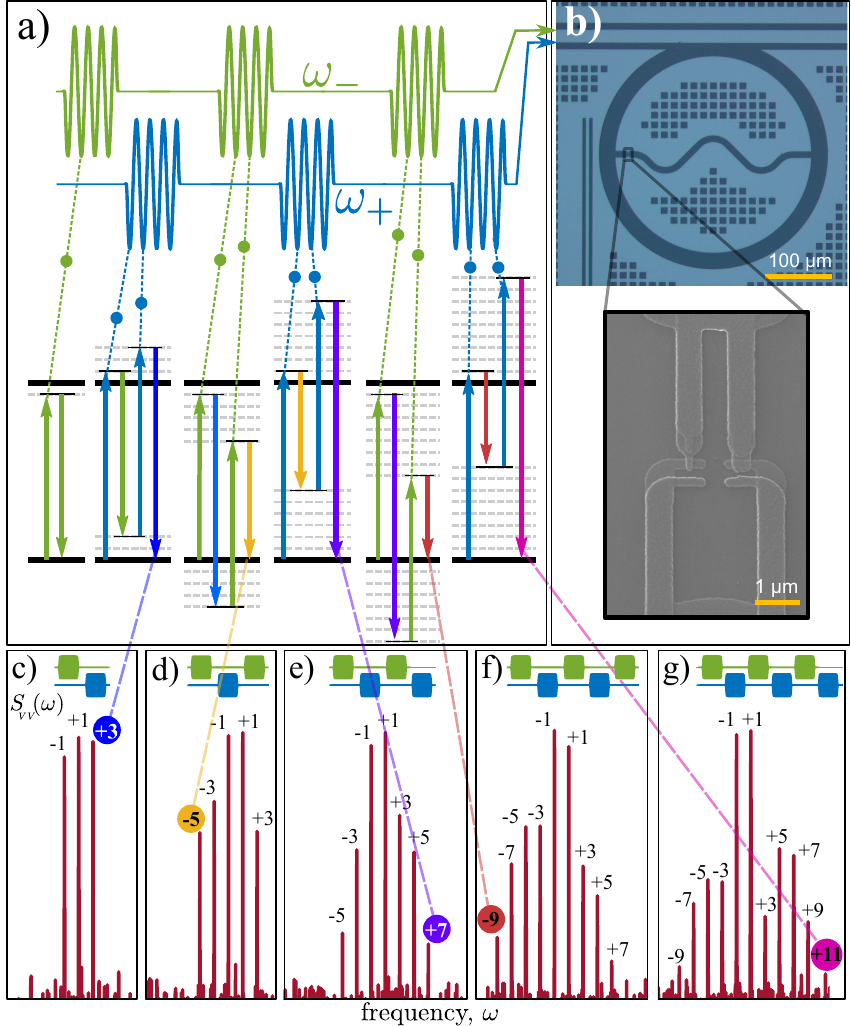}
\caption{\label{fig:phys-descr}a) The sequence of $N$ Rabi pulses scattered on a qubit in the process of QWM. With each additional pulse, a multi-photon process with a pair of photons is emerged. This is illustrated as transitions between dressed levels shifted by $\delta\omega$ from original ones. Therefore the number of elastic peaks is increased with $N$.  b) The SEM image of the transmon qubit. c)-g) The typical scattering spectra measured for $N$ from 2 to 6, respectively. The processes with $2N$ scattered photons are highlighted in color.}
\end{figure}

For the time-domain multiphoton QWM experiment, we utilize a single transmon qubit \cite{Koch2007} as an artificial atom. The qubit consists of a shunting capacitor with the charging energy of $E_C/h = 309$~MHz and an asymmetric SQUID \cite{Hutchings2017} with the total Josephson energy of $E_{J \Sigma}/h = (E_{J1}+E_{J2})/h = 17.0$~GHz and the junction asymmetry $d = 0.5$, giving the upper sweet spot resonance transition at $\omega_q/2\pi = 6.1$~GHz. The qubit is coupled to the central conductor of an on-chip coplanar waveguide with the effective capacitance $C_c = 9.6$~fF, which results in the radiative decay constant  $\Gamma_1/2\pi = 1.64$~MHz in the upper sweet spot. The chip with the qubit is placed in a dilution refrigerator at temperature of 10 mK inside a special holder covered by a magnetic shield. In order to provide thermal equilibrium and allow a heterodyne detection of a scattered signal, coaxial wiring with attenuators, isolators, and HEMT amplifiers is assembled.  The optical and electron images of the device are depicted in Fig.~\ref{fig:phys-descr}. In the spectroscopic measurements, we obtain the power extinction of 99 \% and characterize internal losses and dephasing with the quality factor $Q_i \approx 22000$ defined from circle-fit algorithm \cite{Probst2015}.

We prepare the input pulses with a typical heterodyne up-conversion setup with rf-sources, AWGs (arbitrary waveform generators), and IQ-mixers. As a result, two sequences of short pulses with carrier frequencies $\omega_{\pm}$ are simultaneously propagating near the qubit. Figure~\ref{fig:phys-descr} displays an example of a sequence of 6 pulses in total. The pulse duration is typically 5-10 ns, which guarantees that the qubit keeps some coherence while interacting with every light pulse. We also introduce the gap of 2-4 ns in between pulses with different carriers, ensuring no overlap due to some parasitic reflections or rise and fall effects from AWG. This is crucial because mixing of overlapping pulses on a single two-level system gives an entirely different picture of side peaks \cite{Dmitriev2017}. The repetition period for each sequence is chosen to be 10 $\mu$s or more, which allows the qubit to decay freely in between the pulse sequences so that the initial state before the following pulse sequence is always the ground state with very good precision.

We down-convert the scattered light and digitize both quadratures at an intermediate frequency of 100 MHz to characterize the output field. We use high-frequency ADC with an embedded FPGA, which is programmed to average the identical traces recorded when the trigger is sent from AWG. Since we are interested here in the elastic components, the averaging time is much larger than the period of the sequence. The convenient choice is the time divisible by the period of beats with frequency $2\delta\omega$ because it allows acquiring all the temporal variations of the signal. After recording the complex average field, we make Fourier transformation to get the complex spectrum of the signal analyzed below. 

The lower panels of Fig.~\ref{fig:phys-descr} present a qualitative picture of QWM together with the measured spectra. The leftmost spectrum is obtained for the two pulses. The first pulse is at $\omega_-$ and the second is at $\omega_+$. Only one side component emerges at $\omega_{+3}$ \cite{Dmitriev2017}. Adding a third pulse at $\omega_-$ results in the appearance of two more components at $\omega_{-3} = 2\omega_--\omega_+$ and $\omega_{-5} = 2\omega_--\omega_{+3} = 3\omega_--2\omega_+$. The peak at $\omega_{-5}$ corresponds to the mixing of six photons, which is the highest order of mixing for three pulses. It appears when one photon is taken from the first pulse, a pair of photons from a second one, one more pair from the third one, and an extra photon is emitted at $\omega_{-5}$. With the fourth pulse at $\omega_+$, two more peaks appear at $\omega_{+5}$ and $\omega_{+7}$ in a similar scenario as described for three pulses. To summarize, $N$ applied pulses give rise to all processes up to $2N$-wave mixing.

\begin{figure}[t]
\includegraphics[width=1\linewidth]{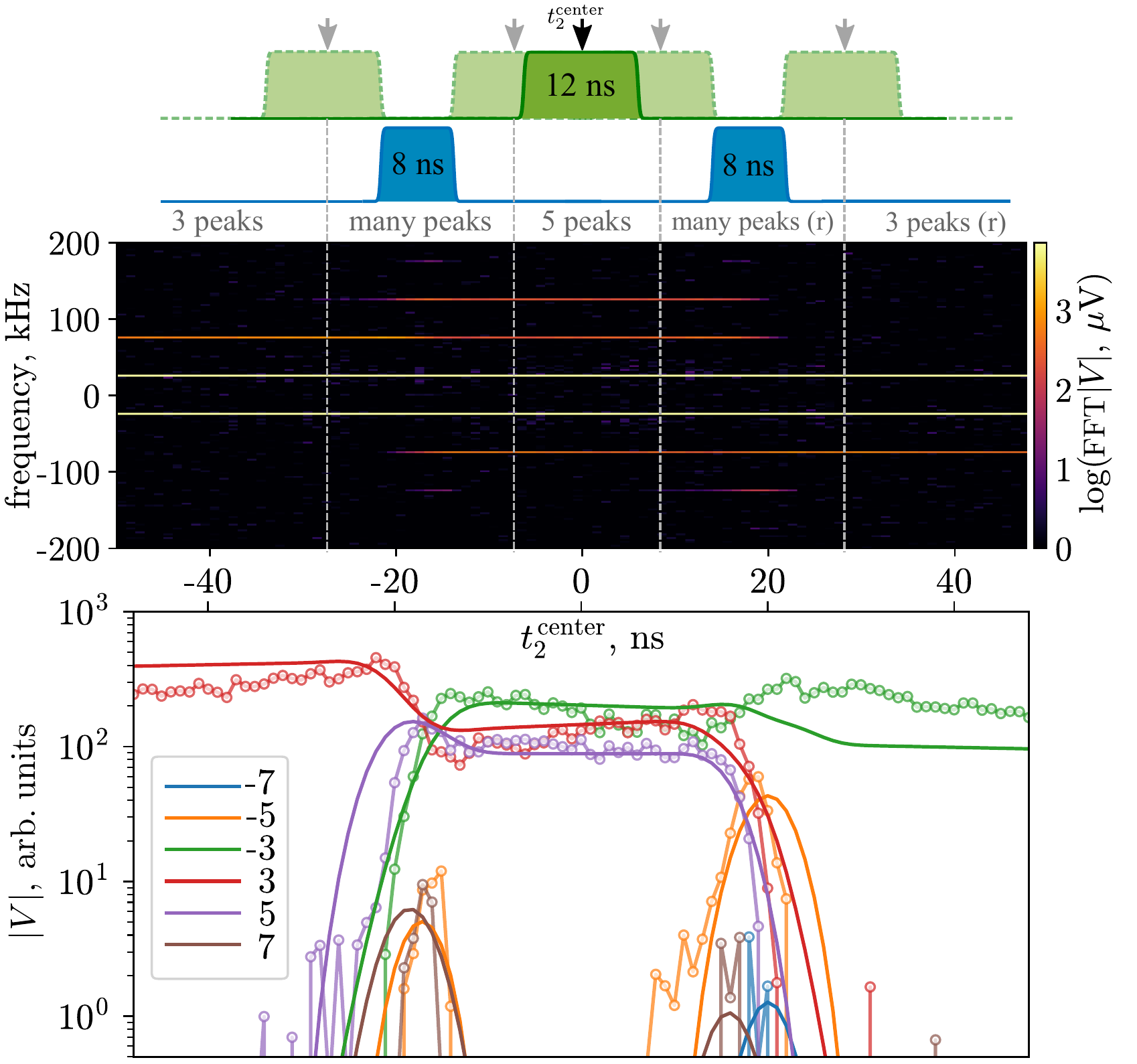}
\caption{\label{fig:shifting}(upper panel) Sketch of sequence of three pulses: blue ones are with $\omega_+$ carrier frequency, and the green is with $\omega_-$. The absolute voltage of side components is plotted as a function of center position of the middle pulse $t^\text{center}_2$. The regions with specific number of components are separated by dashed vertical lines. (bottom panel) The peak amplitudes are rescaled and plotted along with the numerical simulation results. }
\end{figure}

To demonstrate that the observed spectral properties are controllable and may alter by time order of driving pulses, we apply three pulses: two with $\omega_+$ and 8 ns long and one with $\omega_-$ 12 ns long, see Fig.~\ref{fig:shifting}. The position of $\omega_+$ pulses is fixed, while $\omega_-$ pulse is moving. We start with the $\omega_-$ pulse being in front of two pulses at $\omega_+$, and finish when the pulse at $\omega_-$ is after two pulses at $\omega_+$. We observe a consistent picture of several regimes: (i) three peaks are observed when the negatively detuned pulse is either the first or the last and does not overlap with any of $\omega_+$ pulses; (ii) many peaks (7 or more) are observed when there is an overlap between $\omega_+$ and $\omega_-$ pulses, and (iii) five peaks when the pulses are applied one by one as depicted at the top panel of Fig~\ref{fig:shifting}. Therefore, the temporal configuration of driving pulses controls the qualitative properties of QWM coherent spectrum.

\begin{figure*}[t]
\includegraphics[width=1\linewidth]{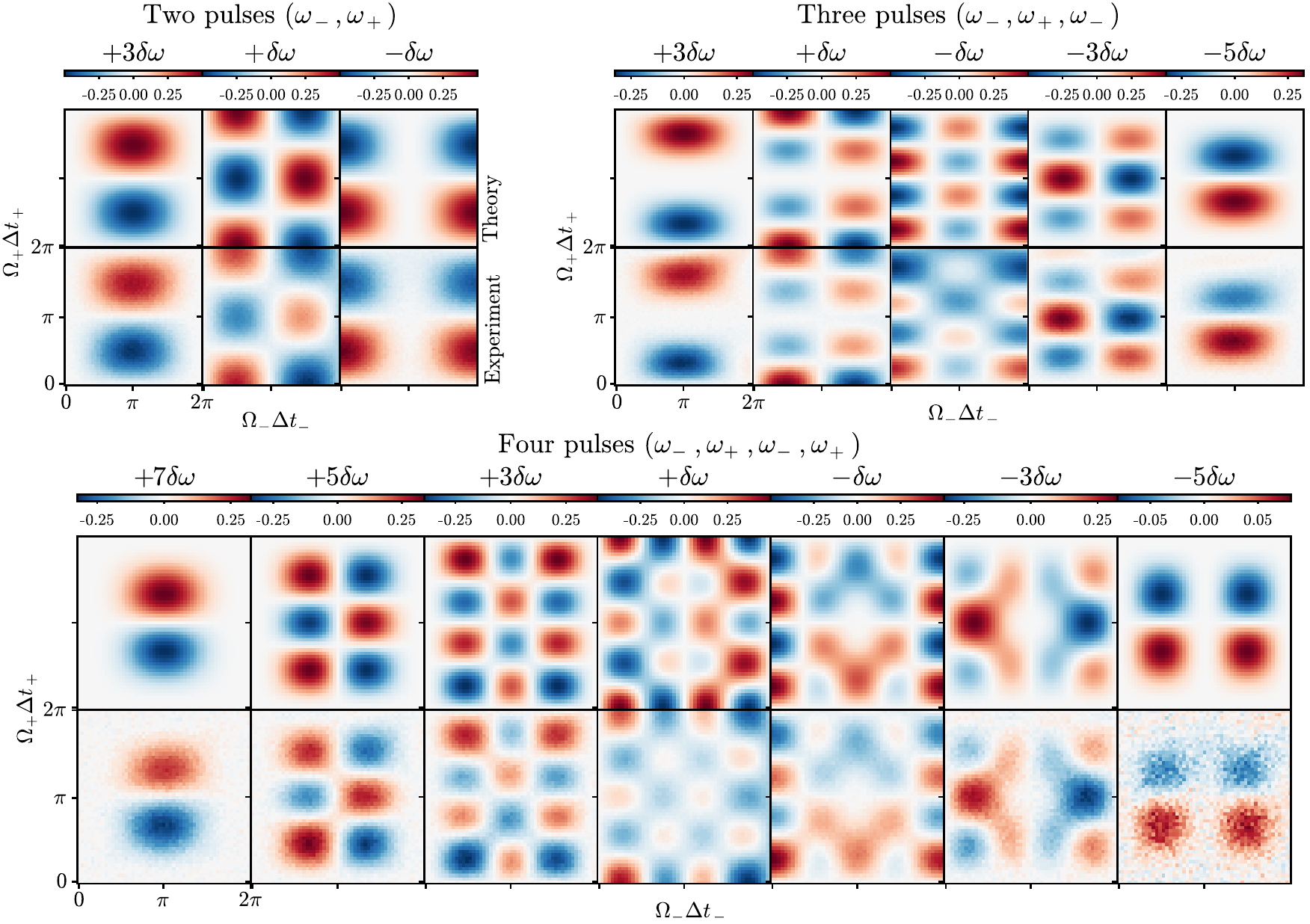}
\caption{\label{fig:2d-pics} The heterodyne-measured complex-valued field of all observed side peaks for sequences of $N$ pulses plotted as functions of $\Omega_+\Delta t_+$ and $\Omega_-\Delta t_-$. The theoretical maps are the result of analytical calculations explained in text.}
\end{figure*}

Next, we characterize quantitative properties of side peaks. For that, we measure how durations $\Delta t_-$, $\Delta t_+$ and amplitudes $\Omega_-$, $\Omega_+$ of pulses affect the intensities of QWM components, implying that similar pulses with the same carrier frequencies rotate qubit states by equal angles. As described earlier, we make complex Fourier transformation of down-converted field, which allows us to extract amplitude and phase of any coherent component.  It is also possible to rotate complex traces such that all the emitted field is concentrated in one quadrature and takes either positive or negative values. We plot color maps of these quadratures versus two arguments $\Omega_-\Delta t_-$, $\Omega_+\Delta t_+$ for each number of pulses $N$. As a result, we obtain several expressive maps demonstrating the rich dynamics of fields, see Fig.~\ref{fig:2d-pics}. We interpret the maps by explicitly calculating the evolution of the qubit under a sequence of pulses. Analytical expressions are given for each spectral component. 
\newline

In order to interpret the observed elastic spectra in terms of multi-photon scattering, we use the framework developed in \cite{Dmitriev2017}. Briefly, we consider two modes of the field $c_-, c_+$ with the frequencies $\omega_+, \omega_-$ interacting with a qubit operators $\sigma^-_{\pm p} \equiv \sigma^- e^{\mp ip\delta t} , \sigma^+_{\pm p} \equiv \sigma^+ e^{\pm ip\delta t}$ which include phase acquired from coherent drive. 

The Hamiltonian in the interaction picture reads
\begin{equation}
    H = ig(c_-^\dag \sigma^-_- - c_- \sigma^+_- + c_+^\dag \sigma^-_+ -  c_+\sigma^+_+).
\end{equation}
To get the field components, we first calculate the following matrix element:
\begin{equation}
    M^{(N)}_p = \braket{\phi_0| \Lambda_N^{\dag}  \sigma^-_p  \Lambda_N | \phi_0},
\end{equation}
where the initial state is $\ket{\phi_0} = \ket{0, \gamma_-, \gamma_+}$, meaning that the scattered field mode is initially in the ground state, and the driving fields are always in coherent states $\ket{\gamma_-}$ and $\ket{\gamma_+}$. The interaction with modes only happens when a corresponding pulse reaches the qubit. The evolution operator $\Lambda_N$ for $N$ applied pulses is defined as 
\begin{equation}
\Lambda_N = \prod_{i=1}^{N} U_{(-1)^i}(\Delta t_i),
\end{equation}
and evolution operators for every single applied pulse with the duration $\Delta t_i$ may approximately be expressed in the case of $\gamma_-, \gamma_+ \gg 1$ as
\begin{equation}
    U_{\pm 1} \approx \cos{\frac{\theta_{\pm}}{2}} + \frac{c^\dag_\pm \sigma^-_\pm - c_\pm \sigma^+_\pm}{\gamma_\pm}\sin{\frac{\theta_{\pm}}{2}}.
\end{equation}
Here we introduced rotation angles $\theta_\pm = \Omega_{\pm} \Delta t_i$ and used $\Omega_\pm = 2g \gamma_{\pm}$. For convenience, we also define the following operators:
\begin{equation}
a^\dag = \alpha c_- \sigma^+_- /\gamma_-, \quad b^\dag = \beta c_+ \sigma^+_+ /\gamma_+
\end{equation}
where $\alpha = \tan\theta_-/2$, $\beta = \tan\theta_+/2$. 

With these expressions, for the case $N\!=\!2$, one obtains:
\begin{widetext}
\begin{equation}
M^{(2)}_p =\left(\cos{\frac{\theta_+}{2}}\cos{\frac{\theta_-}{2}}\right)^{-2} \braket{\phi_0|\Big(
	-\underbrace{\sigma^-_pb^\dag}_{e^{i\delta t}}
	-\underbrace{\sigma^-_pa^\dag }_{e^{-i\delta t}} 
	+\underbrace{a b^\dag \sigma^-_p  a^\dag}_{e^{i\delta t}} 
	+\underbrace{a b^\dag\sigma^-_pb^\dag  }_{e^{3i\delta t}}
	\Big) |\phi_0},
 \label{eq:wideeq}
\end{equation}
\end{widetext}
where under brackets denote the total phase multiplier acquired from operators $a,b,a^\dag,b^\dag$, not counting the phase of $\sigma^-_p$ for a moment. If we now calculate the time-average field at an arbitrary frequency $\omega_{p}$: 
\begin{equation}
V^{(2)}_p = -i\frac{\hbar \Gamma_1}{\mu}\frac{1}{T}\int_0^T M^{(2)}_p \, dt,
\end{equation}
the last term of Eq.~(\ref{eq:wideeq}) will give a non-zero component for $p\!=\!3$, since the phase multiplier from $\sigma^-_{+3} = \sigma^- e^{-3i\delta t}$ cancels out the phase picked from operators of driving modes. Therefore we see that the phases underlined in ~Eq.(\ref{eq:wideeq}) define the frequencies of non-zero elastic components observed when measuring the field spectrum with a low bandwidth.  It explains the leftmost spectrum in Fig.~\ref{fig:phys-descr} which consists of only three components. We also derive how their amplitude depends on rotation angles:
\begin{equation}
V^{(2)}_3 \propto \frac{1}{2}\sin^2 \frac{\theta_+}{2} \cos{\theta_-}
\label{eq: v23}
\end{equation}

It is now straightforward to generalize the calculation for $N>2$ pulses. For three pulses, the analogue of Eq.~(\ref{eq:wideeq}) also expresses $M^{3}_p$ as a sum of terms, and among them there is only operator term $b a^\dag \sigma^-_p a^\dag b a^\dag
$ with acquired phase $e^{-5i\delta t}$, contributing to the peak at $\omega_{-5}$. This peak is highlighted on the second panel from the left in Fig.~\ref{fig:phys-descr}. For the field, we obtain
\begin{equation}
V^{(3)}_{-5} \propto  \frac{1}{2}\sin\theta_-\sin^2\frac{\theta_-}{2}\sin^2\frac{\theta_+}{2}
\label{eq: v35}
\end{equation}
The spectrum of emitted field recorded for different $\theta_-$ and $\theta_+$ allows us to reconstruct how every observable harmonic depends on rotation angles. We then compare these results with the corresponding analytical dependencies, similar to Eqs.~(\ref{eq: v23}) and (\ref{eq: v35}). Both measured and calculated intensities are presented in Fig.~\ref{fig:2d-pics} for $2\le N \le 4$. Supplemental Materials \cite{SuppMat} contain measured results for $N = 5,6$. For all measured components, the theory fits experimental maps very well, even for very high orders. Although, we notice a small disagreement for the emission at the frequency of the last pulse (either $\omega_+$ or $\omega_-$) at large effective angles. The origin of this discrepancy could be in our data processing procedure. In order to restore dependencies for $\omega_+$ and $\omega_-$, we cut out the initial pulses from the digitized traces, replacing them with zero. Therefore, a small part of the emission gets lost. Besides, some part of the last pulse might get distorted due to presence of a small impedance mismatch in the signal line. It may cause a significant effect if the rotation angle of the pulse is large.

Analyzing the patterns in Fig.~\ref{fig:2d-pics}, we can outline several specific features of the observed dependencies. First, we note that for $\theta_-=\theta_+=\pi$, all components are zero for all values of $N$. In this case, each pulse is a $\pi$-pulse; hence the qubit reaches either ground or excited state, does not contain any phase from the pulses and emits incoherently. Second, when the first pulse is negatively detuned, for all values of $N$, $\omega_{ (4p-1)}$-components are zero for $\theta_+ = \pi$, and the maps are anti-symmetrical along the line $\theta_+ = \pi$. Similarly, the maps for $\omega_{(4p-3)}$-components are anti-symmetric relative to $\theta_-=\pi$ and are zero for this angle. It implies that for $\theta_-=\pi$ ($\theta_+=\pi$), there are many values of $\theta_+$ ($\theta_-$), for which emission contains only components at $\omega_{(4p-3)}$ ($\omega_{(4p-1)}$). Similar non-trivial spectral distribution was recently predicted to appear in QWM of a squeezed vacuum with a coherent field \cite{Pogosov2021}. Although, here the origin of missing peaks is due to pulsed drive. 

Our analytical results allow the exact identification of multi-photon contributions into every single mode. As $N$ increases, the higher-order photon processes contribute more significantly to the emission at the frequencies of initial drives and the neighboring frequencies. For example, for $N=3$, the field at $\omega_-$ contains only one term corresponding to a single-photon process (single-photon absorption and emission, or in analogy with higher orders, ``two-wave mixing''), two terms related to the third-order processes (four-wave mixing) and one term connected with the fifth-order process (six-wave mixing). The exact expressions are presented in Supplemental Materials \cite{SuppMat} to this article. Therefore, the interference of all these terms with different orders forms the overall pattern for every component. However, there is always only one term in the evolution responsible for the emission at the leftmost and rightmost peaks frequencies.

\begin{figure}[t]
\includegraphics[width=0.95\linewidth]{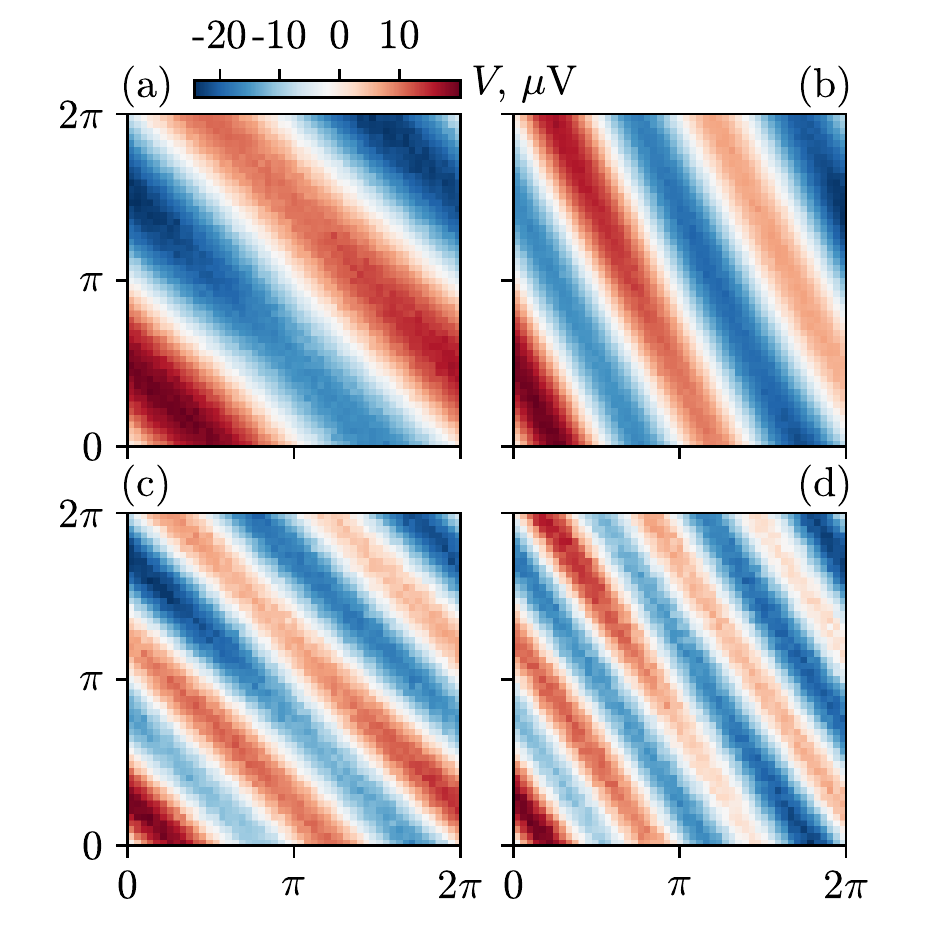}
\caption{\label{fig:sums} The net field obtained as a sum of all measured coherent components, as if it could be emitted for the drives with $\delta\omega = 0$, that is, for a monochromatic drive. From (a) to (d) the map is plotted for N=2,3,4,5 pulses, horizontal axis is $\theta_-$ and vertical is $\theta_+$. The cross-sections are harmonic Rabi oscillations.}
\end{figure}

The observed pictures could be considered as a spectral decomposition of coherent pulses being scattered on a qubit. To illustrate the physical meaning of this decomposition, we can consider the limit $\delta\omega \rightarrow 0$. Our interleaved pulses then turn to continuous monochromatic wave exactly resonant with the qubit. The side peaks now all have the same frequency, and we can simply sum up all the contributions presented in Fig.~\ref{fig:2d-pics} for each sequence, which results in simple Rabi dynamics, with a rotation periods corresponding to total number of pulses, see Fig.~\ref{fig:sums}.

Another observation relates to the efficiency of conversion. For two applied pulses, one can see from Fig.~\ref{fig:2d-pics}a) that if $\theta_- = \pi$ and  $\theta_+=\pi/2$, then all coherent field is emitted into $\omega_{+3}$ mode. We could call this point the maximal conversion point. However, note that there is also incoherently emitted radiation due to inelastic scattering. Correspondingly, for $N>2$ pulses, the distribution of coherent field among side spectral components is more complicated. Nonetheless, there are also rotation angles where the only non-zero component is $\omega_{-3}$.
In summary, we study QWM of light pulses on a single qubit that reveals the intrinsic connection between qubit dynamics and allowed multi-photon processes of elastic scattering. The study of these effects will increase understanding of nonlinear quantum optics with quantum objects playing the role of a scatterer. Recent theoretical efforts show that the wave mixing of classical and quantum fields on a qubit in a waveguide is a good tool for measuring the photon statistics of the quantum signal. However, qubit itself is also a reliable source of non-classical radiation when driven continuously or by short pulses. Non-trivial coherent spectra probably indicate generation of light with more sophisticated statistics than just squeezing or antibunching. Thus, a promising area opens up for further research of microwave optics and photonics.

\begin{acknowledgments}
We wish to acknowledge the support of Russian Science Foundation (grant N 21-42-00025). We thank W. Pogosov for useful discussions.
\end{acknowledgments}

\bibliography{bibliography}

\end{document}


\newcommand{\comment}[1]{}
\renewcommand\thefigure{S\arabic{figure}} 
\renewcommand{\theequation}{S\arabic{equation}}
\title{Supplementary materials for "Time dynamics of multi-photon scattering in a two-level mixer"}

\date{\today}
\keywords{Suggested keywords}
\maketitle

\tableofcontents

\section{\label{fab}Sample fabrication}

The chip with individual qubits coupled to a coplanar waveguide is fabricated with typical three-lithography process. The high-resistance Si wafer was cleaned with the IPA and RCA-1 solution. Surface silicon oxide layer contains many unwanted defects and thereby it is removed by HF after cleaning. Next, the aluminum film of 100 nm is evaporated onto clean wafer with e-beam evaporation machine Plassys MEB-550 in a vacuum better than $10^{-7}$ mBar. The mask creating the pattern for rough structures was created with two-layer stack of resists: 800 nm-thick PMGI resist was covered with 200 nm of positive resist AZ-1505. The stack was baked on hotplate at 160 $^{\circ}$C for 5 minutes and then the mask pattern is exposed by laser maskless lithographer Heidelberg MLA100. The exposed resist is developed with chlorine-based ICP etching, and thereby waveguide and contact pads are formed. Following, the stack for e-beam lithography is formed with Copolymer and ARP-6200.04 resists, which is exposed by electron beam lithographer Crestec-CABL9000 in order to form josephson junctions. After the mask is developed, we perform shadow-angle two-layer evaporation of junctions with the thicknesses of 25 and 40 nm respectively, and then connect junctions to qubit's capacitor with bandages. This requires another lithography and evaporation with removing aluminum oxide formed on top of previous layer by \textit{in situ} Ar ion milling.  The test junctions are fabricated along with qubits, and their resistance is tested with the probe station.
 
 \section{\label{meas}Measurement setup}
 \begin{figure*}[h]
\includegraphics{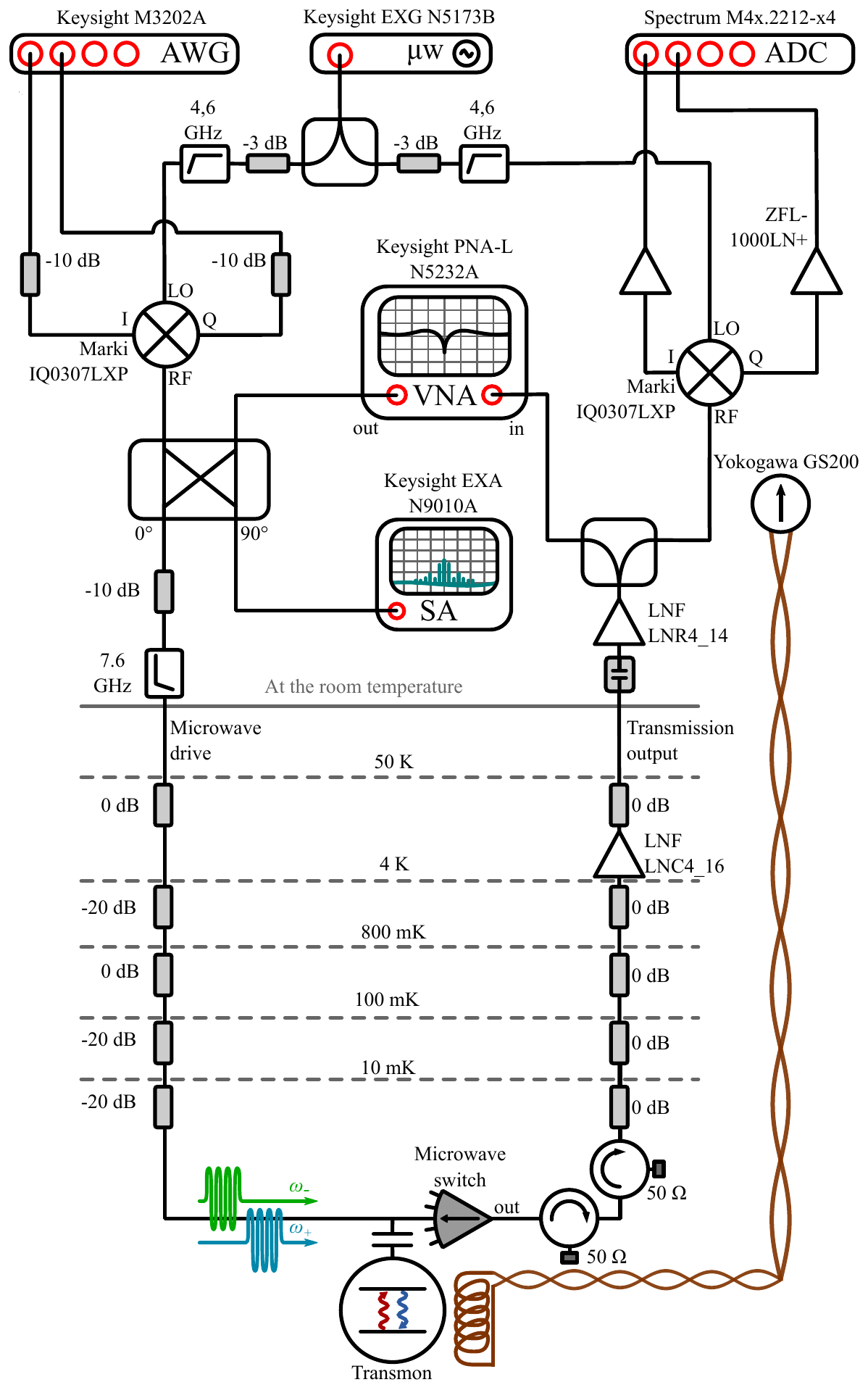}
\caption{\label{fig: scheme}The experimental scheme for measurements of fields scattered by a single qubit in the waveguide.}
\end{figure*}
 The scheme of measurements is presented in Fig.~\ref{fig: scheme}. The room-temperature part includes up-conversion and down-conversion schemes built from commercially available AWGs, mixers, ADCs and rf-generators. For spectroscopic measurements and calibration we also use VNA and SA. The low-temperature part includes input cryogenic coaxial line properly thermalized with attenuators and filters, and output line with amplification chain. Typical solution in our area is to use pair of cryogenic isolators to prevent thermal noise propagation from 4K stage down to sample via output cabling. First of all, our intent is to generate two pulse sequences with close carrier frequencies and different delays, and to do this we use the following scheme of up-conversion. We choose intermediate frequency $\omega_{\text{IF}}$ to be equal to 50-100 MHz, and digitally prepare input waveform, which is the sum of two train sequences of non-overlapping pulses with carriers $\omega_{\text{IF}}+\delta\omega$ and $\omega_{\text{IF}}-\delta\omega$. This waveform is characterized by the period of beats due to small detunings $T_b = 2\pi/2\delta\omega$.  Therefore, the duration of waveforms is conveniently chosen to be the multiple of $T_b$. The individual pulses are much shorter than $T_b$. The time between rising edge of two consecutive pulses with a certain carrier frequency is the fraction of $T_b$, but nevertheless it is much greater than the relaxation lifetime of a qubit. The waveforms are loaded into AWG, the output channels of which are connected to I and Q ports of quadrature mixer. The continuous wave at frequency $\omega_{\text{LO}}$ is given by the rf-generator, output of which is connected to L port of mixer. We calibrate up-conversion by adjusting the phases and offsets of waveforms with carrier frequency of $\omega_{\text{IF}}$ in order to maximize the signal at $\omega_{\text{LO}}+\omega_{\text{IF}}$ and to minimize all other harmonics. 
 
 As for down-conversion part, we exploit the same quadrature mixer. The scattered field undergoes amplification and then comes into $R$-port of the output mixer. The LO signal goes from the same generator as used for up-conversion. Down-converted quadratures from I and Q ports are digitized with the two synchronized channels of ADC, and the acquisition is triggered by AWG.  After moderate averaging (typically millions of waveform periods), we take complex Fourier transform of signal and get the results presented in Fig. 2 and 3 of main text.


\section{\label{sec:spectrum}Qubit Spectroscopy }
\begin{figure}[h] 
\includegraphics[width=0.46\textwidth]{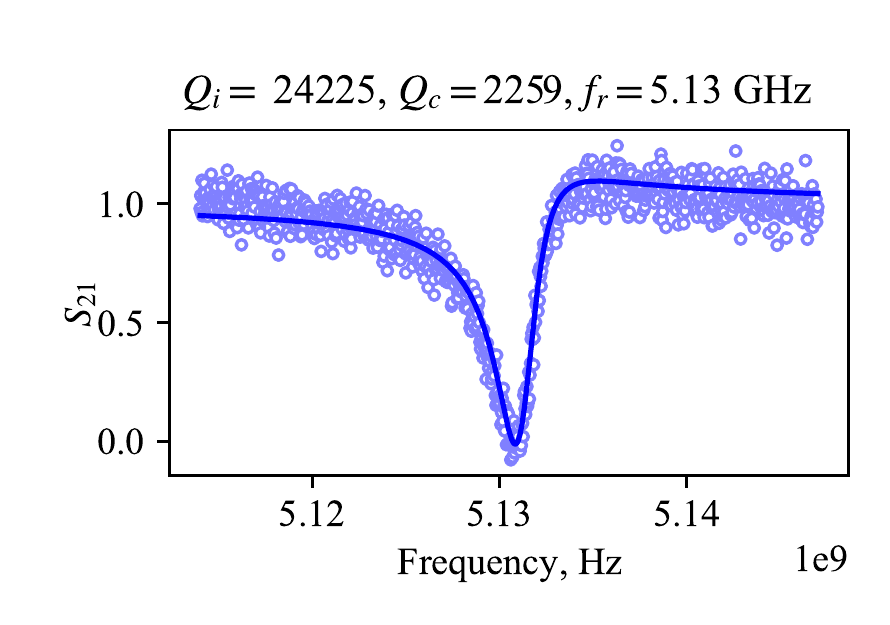} \hfill
\includegraphics[width=0.52\textwidth]{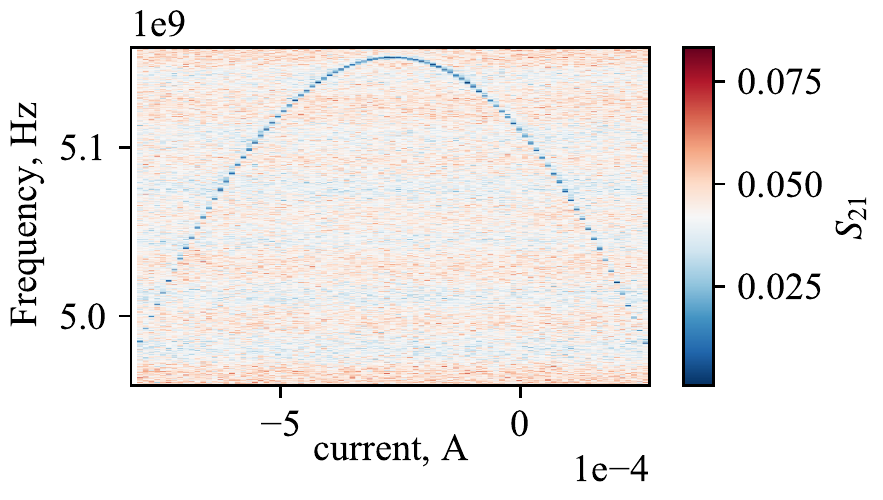} \\
\caption{Single-tone transmission spectroscopy of the qubit in the waveguide. (a) Resonant line of the qubit tuned close to upper sweetspot. Points are data measured with VNA, and line is the best fit to the data. The important fitting parameters are written in the figure title. (b) The current-dependent single tone spectroscopy of the qubit. }
\label{fig: STS}
\end{figure}

The single-tone spectroscopy of the qubit is shown in Fig.~\ref{fig: STS}. We measure the transmission of the signal through the waveguide with the use of VNA. The frequency of stimulus is swept in a wide range, and when it is resonant with the qubit, the dip in transmission is observed, see Fig.~\ref{fig: STS}(a). We normalize the transmission data and fit the qubit line with a circle-fit method. The optimal fit parameters include internal and external (coupling) quality factors $Q_i$ and $Q_c$, respectively, and the frequency of the qubit $f_r$. Sweeping current values and repeating transmission measurement gives us the spectroscopy of the qubit, Fig.~\ref{fig: STS}(b). 

\section{\label{sec: Rabi}Rabi oscillations}
\begin{figure}[h]
\includegraphics[width=0.9\textwidth]{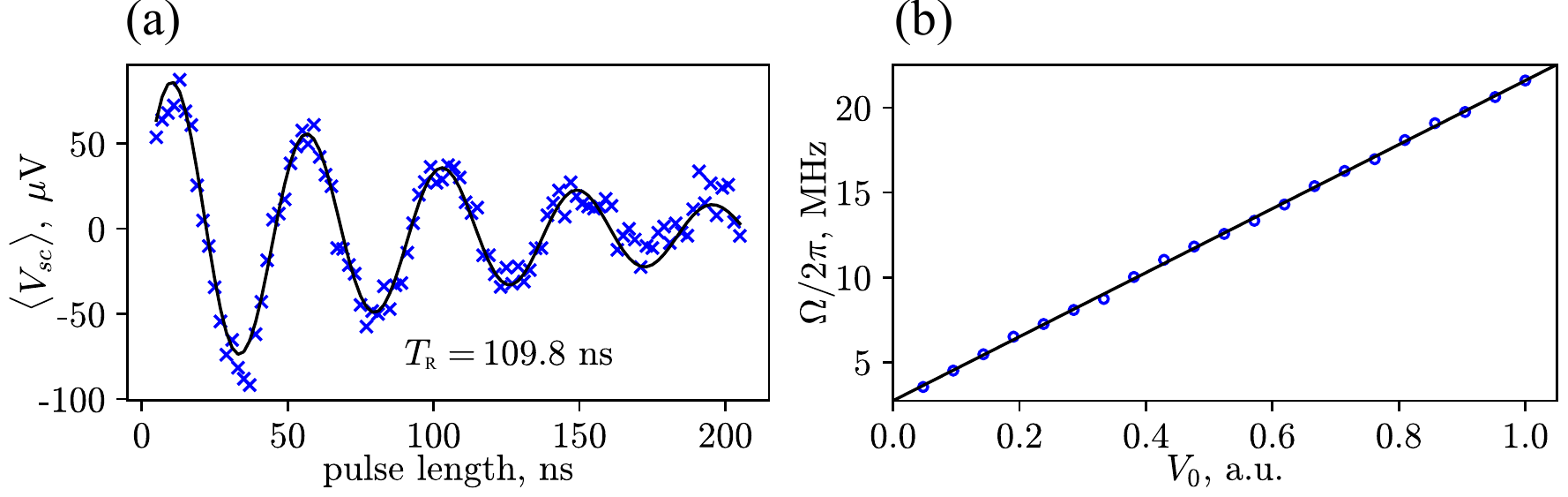}
\caption{Rabi oscillations. (a) A single trace representing qubit dynamics under resonance microwave pulses. Blue crosses show raw data, and the line is optimal fit according to simplified version of Eq.~(\ref{eq: y}). (b) The Rabi frequencies versus voltage amplitude of driving pulses with a linear fit.}
\label{fig: osc}
\end{figure}

With the presented experimental setup we are able to measure the dynamics of the qubit under strong resonant microwave pulses. For this, we direct into a waveguide a sequence of short pulses divided by long intervals. The sequence is formed by our up-conversion scheme. Being resonant with the qubit, the pulse causes the rotational dynamics of the latter, also known as Rabi oscillations. Let us consider the rotation around $x$-axis on the Bloch sphere, then the qubit state evolves in time as:
\begin{align}
\braket{\sigma_x} &= 0, \\ \braket{\sigma_y} &= \frac{2\Omega}{2\Omega^2+\Gamma_1^2}\left[\Gamma_1-e^{-3\Gamma_1 t/4}\left(\Gamma_1\cos\Omega^\prime t + \frac{(4\Omega^2-\Gamma_1^2)}{4\Omega^\prime}\sin \Omega^\prime t\right)\right], \label{eq: y} \\
\braket{\sigma_z} &= -\frac{\Gamma_1}{2\Omega^2+\Gamma_1^2}\left[\Gamma_1+e^{-{3}\Gamma_1 t/4}\left(\frac{3\Omega^2}{2\Omega^\prime}\sin\Omega^\prime t + \frac{2\Omega^2}{\Gamma_1} \cos\Omega^\prime t\right)\right].
\end{align}
Here we suppose that the pure dephasing and non-radiative relaxation are negligible, and that the drive is exactly in resonance with the transition. Also we introduced the generalized Rabi frequency $\Omega^\prime = \sqrt{\Omega^2-\Gamma_1^2/16}$. For measuring the amplitude of the emitted field (which is proportional to $\braket{\sigma_y}$), we down-convert the field emitted right after the driving pulse is gone. After that, we digitize both quadratures the signal at $\omega_{\text{IF}}$ and average over many single traces. Then we apply fast Fourier transform (FFT) in order to get the complex amplitude of field. The single quadrature of field measured for different lengths of driving pulse is presented at Fig.~\ref{fig: osc}(a). 
The fitting of experimental points is done with the help of Eq.~(\ref{eq: y}) and allows to extract the Rabi frequency and the decay time. We repeat this for various driving amplitudes and finally obtain the dependence of $\Omega$ on the voltage amplitude, see Fig.~\ref{fig: osc}(b). 

\section{\label{sec: 5 and 6 pulses}Mixing results for 5 and 6 driving pulses}
\begin{figure}[h]
\includegraphics[width=1\textwidth]{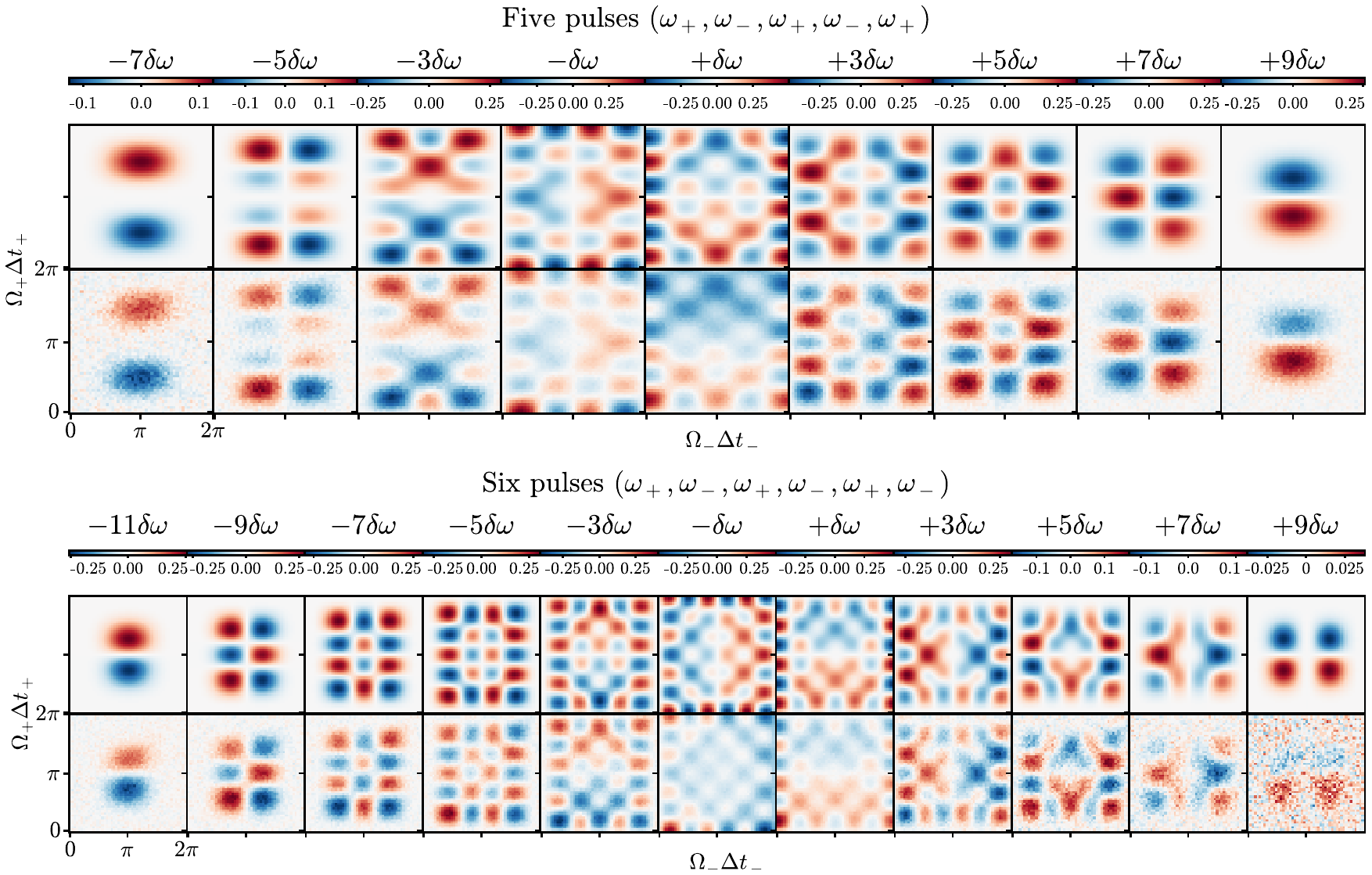}
\caption{The dependence of complex amplitudes of elastically scattered fields on rotation angle of pulses for each side spectral component.}
\label{ed: crop5-6}
\end{figure}

Here we present the extension of results presented in Fig. 3 of main text. The angle dependencies of field amplitudes of side peaks measured for case of five- and six-pulse configuration are presented in Fig.~\ref{ed: crop5-6}. Comparing to the driving sequences with less applied pulses, we can note two features of maps. First, the number of elastic side peaks is increasing: each additional pulse causes the emergence of two extra components.  
Second, for components with frequencies close to initial drives maps are becoming more and more complicated. The reason of this is increasing number of contributions of different orders, whereas for most-detuned components there is only one contribution into furthest components appeared only after the very last pulse.
\section{Analytical expressions for mixing maps}
In this section, we describe in detail how analytical expressions for the mixing components are calculated. To derive the dynamics of $p$-th side spectral components when $N$ pulses are applied,  we calculate the matrix element $M^{(N)}_p$, defined in Eq.~(2) of the main text. It is the expectation value of operator $\sigma^-_p$ on the quantum state of the kind:
\begin{equation}
\ket{\phi(t)} = \Lambda_N\ket{\phi(0)}, 
\end{equation} 
where $\Lambda_N$ is total evolution operator of the system defined in Eq.~(3) of the main text. Operators $\Lambda_N$ can be directly written by multiplying all the evolution operators defined in Eq.~(4), but the resulting expressions are very cumbersome. However, the calculations could be simplified significantly, because we are not interested in the full expression for $\Lambda_N$, but instead looking for $M^{(N)}_p$. In the main text, we introduce operators $a$ and $b$ such that $a^\dag \propto \sigma^+_-$,  $b^\dag \propto \sigma^+_+$. Thus, typically for this kind of calculations, $a\ket{\phi(0)} = b\ket{\phi(0)}= \bra{\phi(0)}a^\dag = \bra{\phi(0)}b^\dag = 0$. As well, any term which contains any of the following multipliers: $(\cdot)(\cdot), (\cdot)^\dag(\cdot)^\dag$, where $(\cdot)=\{a,b\}$  ---  will reduce to zero and give no contribution to $M^{(N)}_p$. This allows to introduce the operator $\Lambda^0_N$ which contains less terms than $\Lambda_N$. Then the simplified expression will be:
\begin{equation}
    M^{(N)}_p = \bra{\phi(0)}(\Lambda^{0}_N)^\dag\sigma^-_p\Lambda^{0}_N\ket{\phi(0)},
    \label{eq: Mn}
\end{equation}

Taking into account Eqs.~(2-5) of main text, we do algebraic computations and express $\Lambda^0_N$ as:
\begin{align}
    \begin{split} \Lambda^{0}_2 &= 1 - a^\dag - b a^\dag - b^\dag, \quad \Lambda_3 =  1 - a a^\dag - a b^\dag - 2 a^\dag + a^\dag b a^\dag - b a^\dag - b^\dag, \end{split} \notag \\
    \begin{split} \Lambda^{0}_4 &= 1 - a a^\dag - a b^\dag - 2 a^\dag + a^\dag b a^\dag - 3 b a^\dag + (b a^\dag)^2 - b b^\dag - 2 b^\dag + b^\dag a a^\dag + b^\dag a b^\dag + b^\dag b a^\dag, \end{split} \notag \\
    \begin{split} \Lambda^{0}_5 &= 1 - 3 a a^\dag + a a^\dag b a^\dag - 3 a b^\dag  +  a b^\dag a a^\dag + (a b^\dag)^2 + a b^\dag b a^\dag - 3 a^\dag + a^\dag a a^\dag + a^\dag a b^\dag + 4 a^\dag b a^\dag - \\ &- (a^\dag b)^2 a^\dag + a^\dag b b^\dag - 3 b a^\dag + (b a^\dag)^2 - b b^\dag - 2 b^\dag + b^\dag a a^\dag + b^\dag a b^\dag + b^\dag b a^\dag, \end{split} \\ 
    \begin{split} \Lambda^{0}_6  &= 1 - 3 a a^\dag + a a^\dag b a^\dag - 3 a b^\dag + a b^\dag a a^\dag + (a b^\dag)^2 + a b^\dag b a^\dag - 3 a^\dag + a^\dag a a^\dag + a^\dag a b^\dag + 4 a^\dag b a^\dag - (a^\dag b)^2 a^\dag + \notag \\ &+ a^\dag b b^\dag - 6 b a^\dag + b a^\dag a a^\dag + b a^\dag a b^\dag + 5 (b a^\dag)^2 - (b a^\dag)^3 + b a^\dag b b^\dag - 3 b b^\dag + b b^\dag a a^\dag + b b^\dag a b^\dag + b b^\dag b a^\dag -\notag \\ &- 3 b^\dag + 4 b^\dag a a^\dag - b^\dag a a^\dag b a^\dag + 4 b^\dag a b^\dag - (b^\dag a)^2 a^\dag - b^\dag (a b^\dag)^2 - b^\dag a b^\dag b a^\dag + 4 b^\dag b a^\dag - b^\dag (b a^\dag)^2 + b^\dag b b^\dag. \end{split}
\end{align}

However, we note that $\Lambda^{0}_N$ in this form still has terms which do not contribute into $M^{(N)}_p$. These are the terms where leftmost operator is $a$ or $b$, and they disappear because products $\sigma^-_pa$ and $\sigma^-_pb$ result in zero during the calculation of (\ref{eq: Mn}). However, for $(\Lambda^{0}_N)^\dag$ the different terms could be removed --- those where the rightmost operator is $a$ or $b$, similarly, because $a\sigma^-_p$ and $b\sigma^-_p$ result in zero. Therefore, (\ref{eq: Mn}) can be now rewritten as:
\begin{equation}
    M^{(N)}_p = \bra{\phi(0)}\Lambda^\leftarrow_N\sigma^-_p\Lambda^{\rightarrow}_N\ket{\phi(0)}.
\end{equation}
Here, operators $\Lambda^{\rightarrow}_N$ and $\Lambda^{\leftarrow}_N$ are obtained from $\Lambda^0_N$ after discarding the proper terms. We come to the following expressions:
\begin{align}
    \begin{split} \Lambda^{\rightarrow}_2 &={} - a^\dag - b^\dag, \quad \Lambda^{\rightarrow}_3 =  - 2 a^\dag + a^\dag b a^\dag - b^\dag, \end{split} \notag \\
    \begin{split} \Lambda^{\rightarrow}_4 &={}  - 2 a^\dag + a^\dag b a^\dag - 2 b^\dag + b^\dag a a^\dag + b^\dag a b^\dag + b^\dag b a^\dag, \end{split} \notag \\
    \begin{split} \Lambda^{\rightarrow}_5 &={}  - 3 a^\dag + a^\dag a a^\dag + a^\dag a b^\dag + 4 a^\dag b a^\dag - (a^\dag b)^2 a^\dag + a^\dag b b^\dag - 2 b^\dag + b^\dag a a^\dag + b^\dag a b^\dag + b^\dag b a^\dag, \end{split}     \label{eq: rightarrow} \\ 
    \begin{split} \Lambda^{\rightarrow}_6  &={} - 3 a^\dag + a^\dag a a^\dag + a^\dag a b^\dag + 4 a^\dag b a^\dag - (a^\dag b)^2 a^\dag + a^\dag b b^\dag - 3 b^\dag + 4 b^\dag a a^\dag - b^\dag a a^\dag b a^\dag + 4 b^\dag a b^\dag - \notag \\ &-{} (b^\dag a)^2 a^\dag - b^\dag (a b^\dag)^2 - b^\dag a b^\dag b a^\dag + 4 b^\dag b a^\dag - b^\dag (b a^\dag)^2 + b^\dag b b^\dag. \end{split}
\end{align}
Similarly, for left-acting operators:
\begin{align}
    \begin{split} \Lambda^{\leftarrow}_2 &= 1 - a b^\dag, \quad \Lambda^{\leftarrow}_3 =  1  - a a^\dag - a b^\dag - b a^\dag, \end{split} \notag \\
    \begin{split} \Lambda^{\leftarrow}_4 &= 1 - a a^\dag - 3 a b^\dag + (a b^\dag)^2 - b a^\dag - b b^\dag, \end{split} \notag \\
    \begin{split} \Lambda^{\leftarrow}_5 &= 1 - 3 a a^\dag  + a a^\dag b a^\dag - 3 a b^\dag + a b^\dag a a^\dag + ( a b^\dag)^2 + a b^\dag b a^\dag - 3 b a^\dag  + a^\dag - b b^\dag, \end{split}     \label{eq: leftarrow}  \\ 
    \begin{split} \Lambda^{\leftarrow}_6  &= 1 - 3 a a^\dag + a a^\dag a b^\dag + a a^\dag b a^\dag  + a a^\dag b b^\dag - 6 a b^\dag + a b^\dag a a^\dag + 5 (a b^\dag)^2 - (a b^\dag)^3 + \notag \\ &+ a b^\dag b a^\dag + a b^\dag b b^\dag - 3 b a^\dag + b a^\dag a b^\dag + (b a^\dag)^2 + b a^\dag b b^\dag - 3 b b^\dag + b b^\dag a b^\dag. \end{split}
\end{align}

With the use of (\ref{eq: rightarrow}) and (\ref{eq: leftarrow}), after trivial algebra we come to the expression for $M^{(2)}_p$, which is presented as Eq.~(6) of the main text. As well, we get expressions for all studied cases, in particular, for 3 and 4 applied pulses:

\begin{multline}
M^{(3)}_p = 
    \cos^{-4}\frac{\theta_-}{2}\cos^{-2}\frac{\theta_+}{2}\bra{\phi(0)}-\underbrace{b a^\dag \sigma^-_p a^\dag b a^\dag}_{\exp{(-5i\delta\omega t)}} 
    + \underbrace{\sigma^-_p a^\dag b a^\dag - a a^\dag \sigma^-_p a^\dag b a^\dag + 2 b                       a^\dag \sigma^-_p a^\dag}_{\exp({-3i\delta\omega t})} \\
    - \underbrace{2 \sigma^-_p a^\dag + 2 a a^\dag \sigma^-_p a^\dag - a b^\dag \sigma^-_p                     a^\dag b a^\dag + b a^\dag \sigma^-_p b^\dag}_{\exp{(-i\delta\omega t)}} \\ 
    - \underbrace{ \sigma^-_p b^\dag + a a^\dag \sigma^-_p b^\dag + 2 a b^\dag \sigma^-_p                      a^\dag}_{\exp{(i\delta \omega t)}} 
    + \underbrace{a b^\dag \sigma^-_p b^\dag}_{\exp{(3i\delta \omega t)}}\ket{\phi(0)},
    \label{eq: M3p}
\end{multline}

\begin{multline}
M^{(4)}_p = \cos^{-4}\frac{\theta_-}{2}\cos^{-4}\frac{\theta_+}{2}\bra{\phi(0)}
             \underbrace{(a b^\dag)^2 \sigma^-_p b^\dag a b^\dag}_{\exp{(7\delta\omega t)}} 
            -\underbrace{ 3 a b^\dag \sigma^-_p b^\dag a b^\dag - 2 (a b^\dag)^2 \sigma^-_p b^\dag + (a b^\dag)^2 \sigma^-_p b^\dag a a^\dag }_{\exp{(5 i\delta\omega t)}}\\
            +\underbrace{\sigma^-_p b^\dag a b^\dag - a a^\dag \sigma^-_p b^\dag a b^\dag + 6 a b^\dag \sigma^-_p b^\dag - 3 a b^\dag \sigma^-_p b^\dag a a^\dag - 2 (a b^\dag)^2 \sigma^-_p a^\dag + (a b^\dag)^2 \sigma^-_p b^\dag b a^\dag - b b^\dag \sigma^-_p b^\dag a b^\dag }_{\exp{(3i\delta\omega t)}} \\
            -\underbrace{ 2 \sigma^-_p b^\dag \!+ \!\sigma^-_p b^\dag a a^\dag\! +\! 2 a a^\dag \sigma^-_p b^\dag\! -\! a a^\dag \sigma^-_p b^\dag a a^\dag\! +\! 6 a b^\dag \sigma^-_p a^\dag\! -\! 3 a b^\dag \sigma^-_p b^\dag b a^\dag\! +\! (a b^\dag)^2 \sigma^-_p a^\dag b a^\dag\! -\! b a^\dag \sigma^-_p b^\dag a b^\dag\! +\! 2 b b^\dag \sigma^-_p b^\dag\! -\! b b^\dag \sigma^-_p b^\dag a a^\dag }_{\exp{(i\delta\omega t)}} \\
            -\underbrace{ 2 \sigma^-_p a^\dag + \sigma^-_p b^\dag b a^\dag + 2 a a^\dag \sigma^-_p a^\dag - a a^\dag \sigma^-_p b^\dag b a^\dag - 3 a b^\dag \sigma^-_p a^\dag b a^\dag + 2 b a^\dag \sigma^-_p b^\dag - b a^\dag \sigma^-_p b^\dag a a^\dag + 2 b b^\dag \sigma^-_p a^\dag - b b^\dag \sigma^-_p b^\dag b a^\dag }_{\exp{(-i\delta\omega t)}} \\
            +\underbrace{\sigma^-_p a^\dag b a^\dag - a a^\dag \sigma^-_p a^\dag b a^\dag + 2 b a^\dag \sigma^-_p a^\dag - b a^\dag \sigma^-_p b^\dag b a^\dag - b b^\dag \sigma^-_p a^\dag b a^\dag }_{\exp{(-3i\delta\omega t)}}
            -\underbrace{b a^\dag \sigma^-_p a^\dag b a^\dag }_{\exp{(-5i\delta\omega t)}}\ket{\phi(0)}.
            \label{eq: M4p}
\end{multline}
Expressions for $N=5$ and $N=6$ are omitted here, but could be written by similar procedure. Finally, substituting operators from Eqs. (5) into Eqs. (\ref{eq: M3p}), (\ref{eq: M4p}), and then using the result in  Eq. (7) of the main text, we come to the final expressions for measured quadrature components of side frequencies. For $N=3$ we have:
\begin{align}
    V^{(3)}_{-5} &= - \sin^{3}{\frac{\theta_-}{2} } \sin^{2}{\frac{\theta_+}{2} } \cos{\frac{\theta_-}{2} }, \qquad
    V^{(3)}_{-3} = \left(2 \cos{\theta_-} + 1\right) \sin^{2}{\frac{\theta_-}{2} } \sin{\frac{\theta_+}{2} } \cos{\frac{\theta_+}{2} }, \notag \\
    V^{(3)}_{-1} &= - \frac{1}{8}\left(3 \cos{\theta_+ } + 1\right) \sin{2 \theta_- }, \qquad V^{(3)}_{1} = \left(1 - 2 \cos{\theta_- }\right) \sin{\frac{\theta_+}{2} } \cos^{2}{\frac{\theta_-}{2} } \cos{\frac{\theta_+}{2} }, \label{eq: V(3)}\\
    V^{(3)}_{3} &= \sin{\frac{\theta_-}{2} } \sin^{2}{\frac{\theta_+}{2}} \cos^{3}{\frac{\theta_-}{2}}. \qquad \quad \notag
\end{align}
For $N=4$, the expressions are:
\begin{align}
    V^{(4)}_{7} &= \frac{ \sin^{2}{\frac{\theta_-}{2} } \sin^{4}{\frac{\theta_+}{2} } \sin{{\theta_-} }}{\cos{\theta_+ } + 1}, \qquad
    V^{(4)}_{5} = - \frac{\left(3 \cos{\theta_- } + 2\right) \sin^{2}{\frac{\theta_-}{2} } \sin^{3}{\frac{\theta_+}{2} }}{\cos{\frac{\theta_+}{2} }}, \notag\\
    V^{(4)}_{3} &= \frac{\left(15 \cos{\theta_- } \cos{\theta_+ } + 11 \cos{\theta_- } + \cos{\theta_+ } + 1\right) \sin{\theta_-} \sin^{2}{\frac{\theta_+}{2} }}{4 \left(\cos{\theta_+ } + 1\right)} \notag\\
    V^{(4)}_{1} &= \left[\sin^{2}{\theta_- }\left(5  \cos{\theta_+ } + 1 \right)   -\cos{\theta_+ }\left(\cos{\theta_- } + 3\right)  \right] \tan{\frac{\theta_+}{2} } \label{eq: V(4)}\\
    V^{(4)}_{-1} &= \frac{ \sin{\theta_-} }{8 \left(\cos{\theta_+} + 1\right)}\left[\left(1 - \cos{\theta_+}\right)^{2}\left(- 15  \cos{\theta_-} - 1\right) + \cos\theta_-\left(20  - 36\cos\theta_+\right)\right] \notag\\
    V^{(4)}_{-3} &= \left(\frac{\cos{\theta_- }}{2}\left(3 \cos{\theta_+ } + 1\right) + \cos{\theta_+ }\right) \sin^{2}{\frac{\theta_-}{2} } \tan{\frac{\theta_+}{2} } \notag \\
    V^{(4)}_{-5} &= - \sin^{3}{\frac{\theta_-}{2} } \sin^{2}{\frac{\theta_+}{2} } \cos{\frac{\theta_-}{2} }. \notag \\
\end{align}
The dependencies (\ref{eq: V(3)}) and (\ref{eq: V(4)}) are plotted in Fig.~3 of main text, and the good agreement with experimentally measured angle dependencies is observed.

\section{Numerical simulation of mixing}
In this section we disclose the technical details of numerical calculations exploited to obtain side spectral components, which emerge as a result of wave mixing between ultrashort bichromatic pulses.
\begin{figure}[h]
    \centering
    \includegraphics[width=1\textwidth]{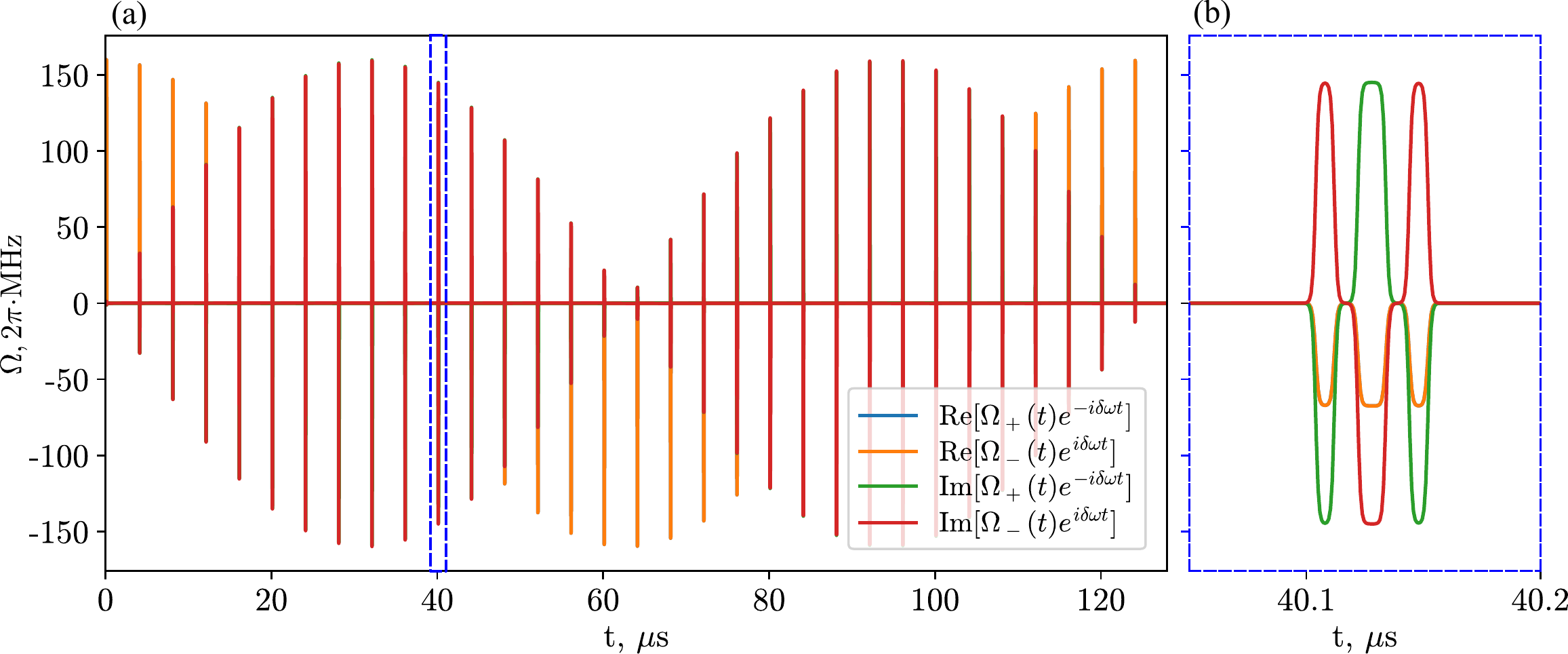}
    \caption{The numerically generated sequence of driving pulses. (a) The one of pulse sequences used to generate lines at Fig. 2 of the main text. The qubit is driven by ultra-short pulses with the length of 8-12 ns, and the period of pulses is 40 us. The slowly varying envelope is from exponential multipliers $e^{\pm \delta\omega t}$, with $\delta\omega/2\pi=50$ kHz. (b) the magnified view of three pulses at $\omega_-, \omega_+, \omega_-$ with the durations of 8,12 and 8 ns. The padding is 8 ns. The sequence is updated at each step, and the relative delay of the middle pulse is swept from -50 to 50 ns with respect to the position plotted in (b).  }
    \label{fig: pulses}
\end{figure}
The numerical simulation is done for different configurations of bichromatic pulsed drive. In previous section we considered the process of analytical derivation of side spectral components especially for non-overlapping pulses. In that relatively simple case, the qubit interacts only with one mode at a time, and never with two modes simultaneously. This allows us to calculate total evolution by multiplying time-dependent evolution operators of single mode Jaynes-Cummings model, which come one by one from distinct pulses. The expression of this evolution operator is commonly known and may be found, for example in \cite{schleich}, Ch. 15. However, for the case of overlapping red- and blue-detuned pulses, the similar calculation becomes less straightforward \cite{Dmitriev2017}, because there is no analytical solution for the atom driven by two detuned continuous waves. Particularly, there is a Bessel function dependencies of side components given in \cite{Dmitriev2017}, but they are only correct if evolution starts from ground state, which is not the case for partially overlapped pulses. For example, if we have two overlapped pulses, then the atom is, first of all, is rotating by a first single mode driving pulse, and already afterwards it evolves under two drives, from the start moment of the second pulse. Therefore, numerical calculations are convenient to demonstrate mixing spectrum for more complicated sequence of driving pulses. 

We numerically simulate the system evolution under time-dependent Hamiltonian under rotating wave approximation:
\begin{equation}
  H_{\text{RWA}}(t)=  \left[\begin{matrix}0 & \Omega_+(t) e^{- i \delta\omega t} + \Omega_-(t) e^{i \delta\omega t}\\\Omega_+(t) e^{i \delta\omega t} + \Omega_-(t) e^{- i \delta\omega t} & 0\end{matrix}\right],
\end{equation}
where we kept only slowly varying terms from both drives. We employ open-source \verb|QuTiP| package to construct the short pulses with envelopes $\Omega_+(t)$ and $\Omega_-(t)$ which are plotted at Fig. \ref{fig: pulses} for three driving pulses. With this, we construct Lindblad master equation with radiational relaxation and then use \verb|mesolve| function to calculate $\braket{\sigma_-(t)}$.
\begin{figure}
    \centering
    \includegraphics[width=0.75\textwidth]{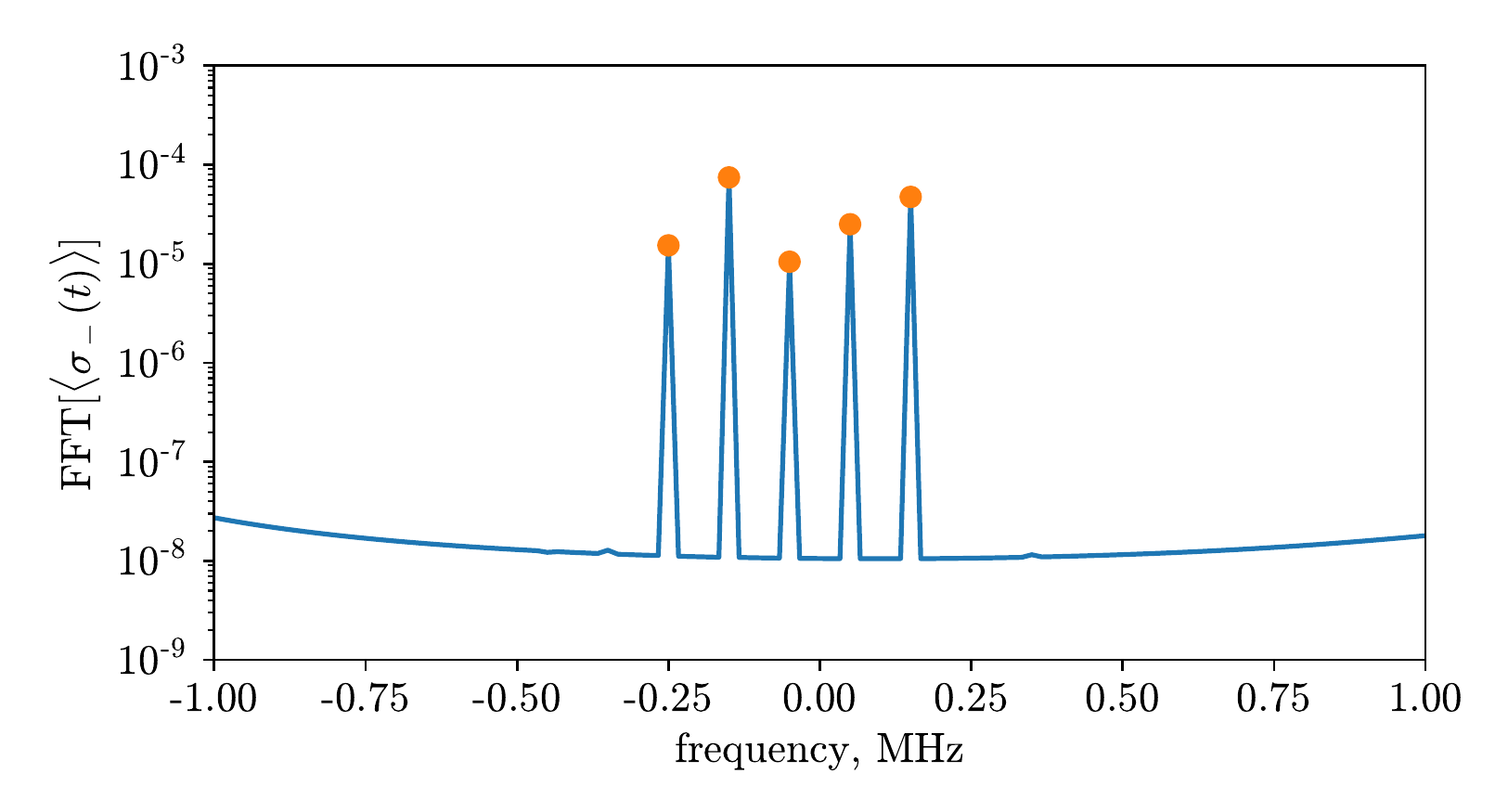}
    \caption{Fourier transform of $\braket{\sigma_-(t)}$}
    \label{fig: fft_spec}
\end{figure}

\begin{figure}
    \includegraphics[width=.495\textwidth]{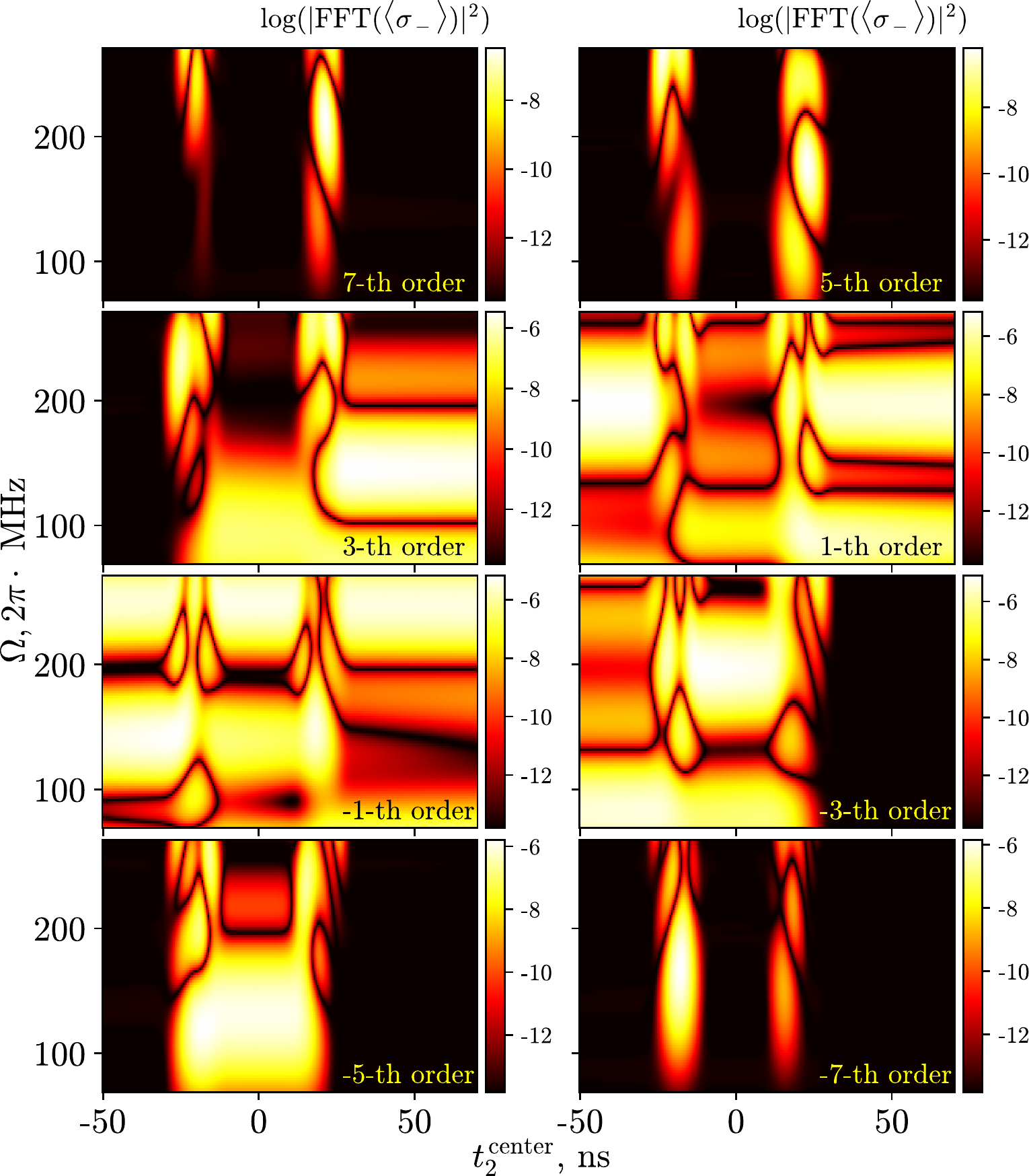} \hfill
    \includegraphics[width=.495\textwidth]{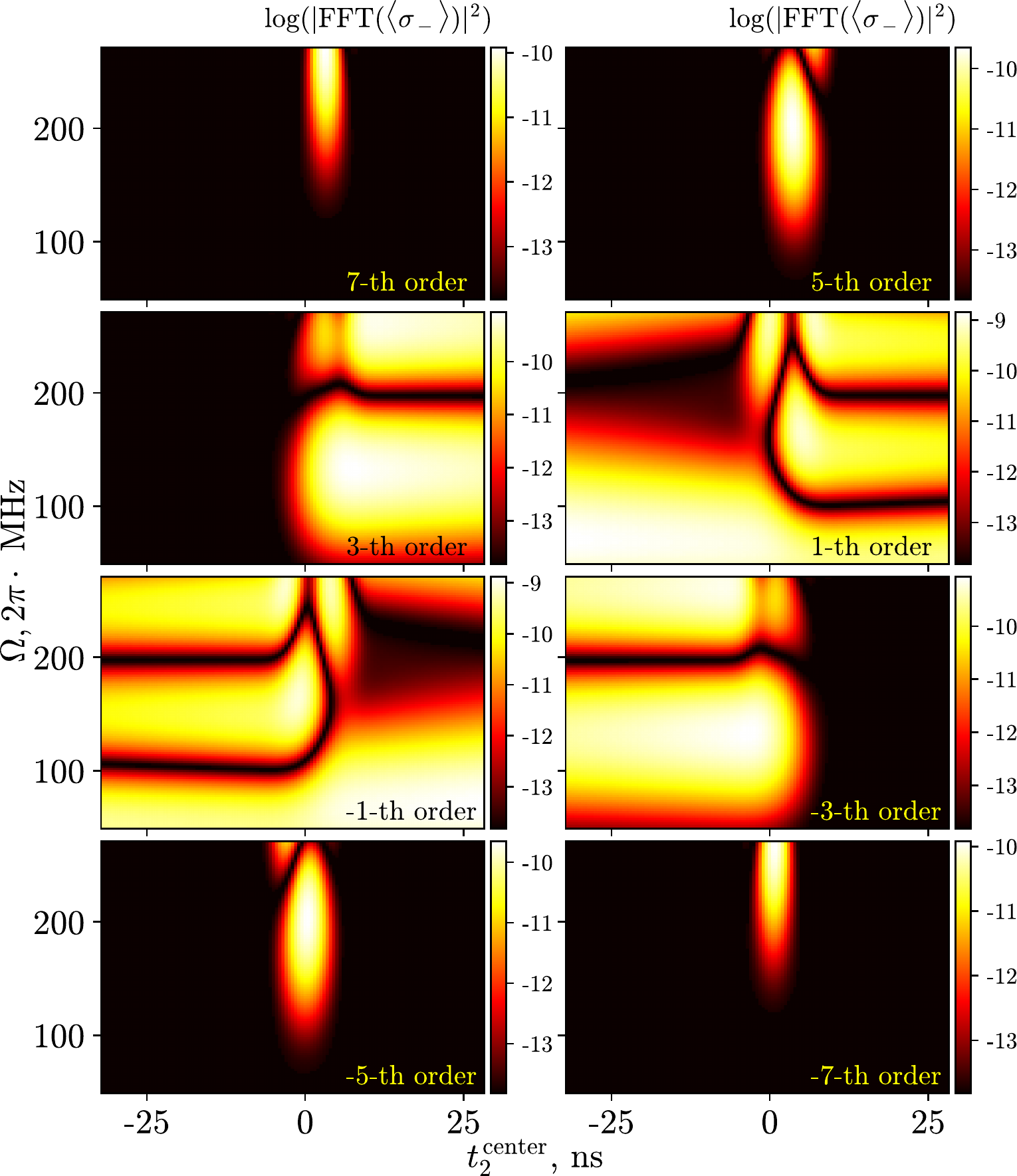} \\
    \parbox[t]{0.47\textwidth}{\caption{Scattering of three-pulse sequence shown at Fig.~\ref{fig: pulses}. Intensities of elastic components are plotted as a function of effective Rabi amplitude $\Omega$ and the middle point $t_2^{\text{center}}$ of the second pulse at $\omega_+$ shifted with respect to another pair of pulses at $\omega_-$. }\label{fig: num_maps_3}} \hfill
    \parbox[t]{0.47\textwidth}{\caption{Scattering of two-pulse sequence. Intensities of elastic components plotted as a function of effective Rabi amplitude of pulses $\Omega$ and the middle point $t_2^{\text{center}}$ of the second pulse at $\omega_+$ shifted with respect to another single pulse at $\omega_-$. }\label{fig: num_maps_2}}
\end{figure}

The Fourier transform of expectation value is presented in Fig.~\ref{fig: fft_spec}. We can distinctly observe three sidebands, two of which are red-shifted and one is blue- shifted. This corresponds to the qualitative picture of multi-photon interaction which we discuss in the main text. However, as we described earlier, the spectrum is more complicated for overlapping pulses. We illustrate the variety of spectral properties by performing above-described simulation for various values of the middle pulse delay and in wide range of Rabi drives $\Omega_+=\Omega_-=\Omega$. The results are presented at Fig.~\ref{fig: num_maps_3}. The all-dark regions for components $\omega_{-7}, \omega_{7}$ and $\omega_5$ are observed for all range of $\Omega$ when the middle pulse (at $\omega_+$) is either very first, very last, or in between other two pulses (at $\omega_-$). This regions correspond to $t_2^{\text{center}}<-28$ ns, $-8<t_2^{\text{center}}<8$ ns and $t_2^{\text{center}}>28$ ns. 
Similarly, we make the simulations for a sequence of two pulses, the result of which also exhibits dark regions where the corresponding mixing processes are prohibited, see~Fig.~\ref{fig: num_maps_2}. As expected, the maps are symmetric with respect to delay and frequency inversion: $\sigma^-_p(\Omega, t_2^{\text{center}}) = \sigma^-_{-p}(\Omega, -t_2^{\text{center}})$.
These results confirm that the missing peaks at Fig. 1 and 2 of main text are not only disappear at specific Rabi angles of driving pulses, but are absent for any amplitudes.


\bibliography{supplement}